# CT Image Reconstruction in a Low Dimensional Manifold

Wenxiang Cong[1], Ge Wang[1], Qingsong Yang[1], Jiang Hsieh[3], Jia Li[2], Rongjie Lai[2]
[1]Biomedical Imaging Center, Department of Biomedical Engineering,
[2]Department of Mathematical sciences
Rensselaer Polytechnic Institute, Troy, NY 12180
[3]GE Healthcare Technologies, Waukesha, WI 53188

**Abstract:** Regularization methods are commonly used in X-ray CT image reconstruction. Different regularization methods reflect the characterization of different prior knowledge of images. In a recent work, a new regularization method called a low-dimensional manifold model (LDMM) is investigated to characterize the low-dimensional patch manifold structure of natural images, where the manifold dimensionality characterizes structural information of an image. In this paper, we propose a CT image reconstruction method based on the prior knowledge of the low-dimensional manifold of CT image. Using the clinical raw projection data from GE clinic, we conduct comparisons for the CT image reconstruction among the proposed method, the simultaneous algebraic reconstruction technique (SART) with the total variation (TV) regularization, and the filtered back projection (FBP) method. Results show that the proposed method can successfully recover structural details of an imaging object, and achieve higher spatial and contrast resolution of the reconstructed image than counterparts of FBP and SART with TV.

## 1. Introduction

X-ray computed tomography (CT) is a major imaging modality in medical, security, and industrial applications. The filtered back-projection (FBP) is an efficient and robust method for x-ray CT image reconstruction [1], but it generates strong image noise and artifacts in the cases of low-dose or incomplete datasets. Extensive efforts have been made to improve image quality for practical purposes [2-4]. Iterative methods incorporate prior information of images, and offer distinct advantages over the analytic methods in cases of noisy and few-view data. The statistical iterative methods model the statistics of photons to improve the reconstructed image quality from the low-dose acquisitions [4, 5]. Recently, the compressive sensing (CS) approach [6, 7] is applied for the image reconstruction from less measurements than that required by the Nyquist-Shannon sampling theorem. Based on the CS theory, image reconstruction algorithms were developed for various problems of CT image reconstruction for improving image quality and reducing radiation dose, such as total variation (TV) regularization [3, 5], nonlocal mean (NLM) [2, 8], dictionary learning (DL) [9], prior image constrained compressed sensing (PICCS) [10] , and prior rank and sparsity model (PRISM)-based image reconstruction [11]. TV is a typical sparse transform for an image, and is a popular regularization form for image reconstruction due to its ability to preserve image edges. However, it is effective only for reconstruction of piecewise constant images and would over-smooth textured regions, which may sacrifice important details. NLM exploits a high degree of redundancy of an image for de-noising [8]. The similarity is derived from intensity differences between neighboring patches of pixels or voxels. A non-linear filter can be used to reduce image noise by updating each pixel value with a weighted average of its neighbors according to the similarity of involved patches. DL builds adaptive sparse representation elements from a training set of images, and utilizes domain knowledge at a deeper level [9]. The dictionary tends to capture local image features effectively and helps image denoising and feature inference. However, the structural differences between a true image and training images could affect the image reconstruction quality. PICSS regularizes image reconstruction with a prior image instead of image patches [10]. PRISM combines sparsity and low rank expectations of an image. All these methods were reported with various degrees of success but no perfect solution exists that is sufficiently accurate and robust, and further improvement in image quality remains a popular topic.

The idea of the proposed X-ray CT image reconstruction model is inspired by a recent method called the low-dimensional manifold model (LDMM) [12, 13]. Using the image patches discussed in nonlocal methods [13], the LDMM interprets image patches as a point cloud sampled in a low-dimensional manifold embedded in a high dimensional ambient space, which provides a new way of regularization by minimizing the dimension of the corresponding image patch manifold. This can be explained as a natural extension of the idea of low-rank regularization for linear objects to data with more complicated structures. Moreover, the authors in [12] elegantly find that the point-wisely defined manifold dimension can be computed as a Dirichlet energy of the coordinate functions on the manifold, whose corresponding boundary value problem can be further solved by a point integral method proposed in [14]. The LDMM performs very well in image imprinting and super-resolution. In this paper, the regularization method based on LDMM is proposed for CT image reconstruction. The patch manifold of images is generally a low dimensional structure, and yet accommodates rich structural information [13]. Using the Bregman iteration [15], the proposed reconstruction model can be iteratively solved by a sequence of soft thresholding, Possion equations provided by the Laplace-Beltrami operator over a point cloud using the point integral method [12], and updating the patch manifold structure by renewing the K-nearest neighborhood.

The rest of the paper is organized as follows. In section 2, we provide a detailed description for the proposed X-ray CT image reconstruction model based on LDMM. A Numerical algorithm is also designed based on Bregman iteration. In section 3, we perform the image reconstruction for the clinical raw projection data from GE Clinic using the proposed LDMM-based reconstruction method. In addition, we also conduct reconstruction comparisons with results obtained from FBP and SART with TV. Our numerical results validate the effectiveness of the proposed method. After that, we conclude the paper in the last section.

## 2. Image Reconstruction Method

In this section, we first review the statistical model of x-ray CT imaging. After that, we will discuss the proposed model of CT image reconstruction based on LDMM and its numerical algorithm.

## 2.1. Statistical Model for X-ray CT imaging

In x-ray CT imaging, the number $\xi$ of x-ray photons recorded by a detector element is a random variable, which obeys a Poisson distribution [1]:

$$p(\xi = y_i) = \frac{(\bar{y}_i)^{y_i}}{y_i!} \exp(-\bar{y}_i). \tag{1}$$

The expectation value of x-ray photons along a path $l$ from x-ray source to *i-th* detector element obeys Beer-Lambert law:

$$\bar{y}_i = b_i \exp\left(-\int_l \mu(\vec{r}) dl\right), \tag{2}$$

where $b_i$ is the number of x-ray photons detected by *i-th* detector element in the blank scanning (without any object in the beam path), and $\mu(\vec{r})$ is the linear attenuation coefficient of the object. To implement the numerical computation, Eq. (2) can be discretized as,

$$\bar{y}_i = b_i \exp(-A_i \mu) \tag{3}$$

where $\mu$ is a vector composed of pixel values on image of linear attenuation coefficients, and $A_i$ is the weighting coefficients of the pixel values on i-th beam path. Since data are independent between detectors, the likelihood function for x-ray photons probability distribution on detectors is,

$$P(Y|\mu) = \prod_{i=1}^{N} \frac{(\bar{y}_i)^{y_i}}{y_i!} \exp(-\bar{y}_i), \tag{4}$$

where $Y = (y_1, y_2, \cdots, y_N)^T$. According to the Bayesian rule: $P(\mu|Y)P(Y) = P(Y|\mu)P(\mu)$, the image reconstruction task can be implemented by maximizing a posteriori (MAP) distribution $P(\mu|Y)$ [5, 16]. From the monotonic property of the natural logarithm, the image reconstruction can be reduced to following minimization problem [5]:

$$\mu = \arg\min\left[\sum_{i=1}^{N}(\bar{y}_i - y_i \log(\bar{y}_i)) + R(\mu)\right], \tag{5}$$

where $R(\mu) = -\ln(P(\mu))$ is a regularization term expressing the prior knowledge about the attenuation image $\mu$, and $N$ is the total number of x-ray beam paths. In the context, we propose to use the low-dimension of an image as prior knowledge to conduct image reconstruction, which is discussed in the next Section. After inserting Eq. (3) in Eq. (5), a second-order approximation is applied to simplify the complicated optimization to a quadratic optimization:

$$\mu = \arg\min \sum_{i=1}^{N}\left(\frac{b_i}{2}(A_i\mu - y_i)^2 + R(\mu)\right) \tag{6}$$

## 2.2. Image Reconstruction algorithm using LDMM

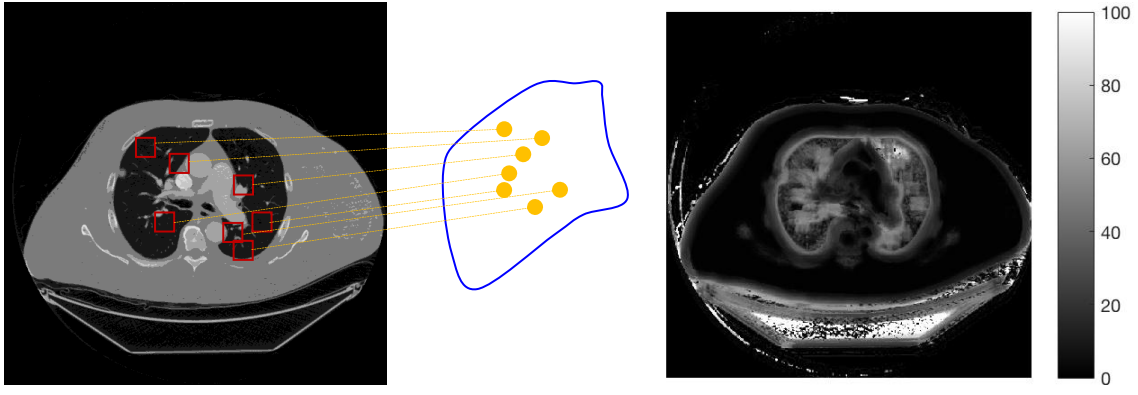

Fig.1. The patch manifold of a CT image (left) and the corresponding dimension function of the patch manifold with patch size 16X16 (right).

Let $I$ denote an image contained $m \times n$ pixels: $I = \{I(i,j) | 1 \leq i \leq m, 1 \leq j \leq n\}$, and $P_s(I)$ denotes a patch of image $I$, which is a sub-image of $I$ with size of $2s_1 \times 2s_2$, $P_s(I) = \{I(i,j) | i_0 - s_1 \leq i < i_0 + s_1, j_0 - s_2 \leq j < j_0 + s_2\}$, here $(i_0, j_0)$ is the central coordinates of the patch. An image is decomposed into a set of patches. These patches can be overlapping or nonoverlapping. Let $P(I)$ denotes all patch set such that the union of the patch set covers the whole image, for example $\Theta = \{1, s_1+1, 2s_1+1, \cdots, m\} \times \{1, s_2+1, 2s_2+1, \cdots, n\}$ is an index set of the patch. $P(I)$ can be also seen as a point set in $R^d$ with a dimension of $d = 2s_1 \times 2s_2$. $P(I)$ samples a low dimensional manifold $M(I)$ embedded in $R^d$, which is called the patch manifold of $I$ as shown in the left image of Fig. 1. The patch manifold is low dimensional for many natural images [13]. In fact, for X-ray CT images, this low-dimensional structure of the patch manifold is also true. As an example illustrated in the right image of Fig. 1, we construct a patch manifold of a CT image using patching size 16X16. This leads to a point cloud, which point-wise dimension is color-coded on the image. Based on this assumption, one natural regularization term is defined as the dimension of the patch manifold to seeking detail structure information for the image reconstruction. This method recovers the CT image such that the dimension of its patch manifold is as small as possible. Therefore, the optimization model Eq. (6) is reformulated for the measurement data fidelity and the manifold dimensional quantification:

$$\boldsymbol{\mu} = \arg\min\left[\frac{\lambda}{2}\sum_{i=1}^{N} b_i(A_i\boldsymbol{\mu} - y_i)^2 + \beta \dim(\mathrm{M}(\boldsymbol{\mu}))\right] \qquad (7)$$

where $\dim(\mathrm{M}(\boldsymbol{\mu}))$ denotes the dimensionality of the patch manifold $\mathrm{M}(\boldsymbol{\mu})$ of an image $\boldsymbol{\mu}$. With differential geometry, the dimensionality of the patch manifold can be calculated by the coordinate function [12],

$$\dim(\mathrm{M}(\boldsymbol{\mu})) = \sum_{i=1}^{d} |\nabla_M \alpha_i(x)|^2 \qquad (8)$$

where $\alpha_i$ is the embedding coordinate function defined by $\alpha_i(x) = x_i$, $i = 1, 2, \cdots, d$, for any $x = (x_1, x_2, \cdots, x_d) \in M \subset R^d$.

Combining Eqs. (7) and (8), we obtain

$$\min_{\mu, M} \left[ \frac{\lambda}{2} \sum_{i=1}^{N} b_i (A_i \mu - y_i)^2 + \beta \sum_{i=1}^{d} |\nabla_M \alpha_i|^2 \right], \quad s.t. \quad P(\mu) \subset M, \tag{9}$$

where $M$ is a manifold, and $P(\mu)$ is the patch set. The optimization (9) can be solved by alternating direction iteration. Given an initial image $\mu$, the manifolds $M$ is established. Then optimization (9) is implemented to update the image $\mu$. From the reconstructed image $\mu$, the manifold is further updated, and image reconstruction is performed. This process is repeated until convergence of iterative procedure. The computation of manifold from an image is direct and easy. Given the manifold $M$, the optimization problem Eq. (9) can be solved to compute the coordinate functions $\alpha_i$ ($i = 1, 2, \cdots, d$) and update the image $\mu$ using the split Bregman iteration [15].

$$\begin{cases} (\alpha_1, \alpha_2, \cdots, \alpha_d) = \arg\min \left[ \sum_{i=1}^{d} |\nabla_M \alpha_i|^2 + \beta \|\alpha(P(\mu)) - P(\mu) + Q\|^2 \right], & (a) \\ \mu = \arg\min \left[ \frac{\lambda}{2} \sum_{i=1}^{N} b_i (A_i \mu - y_i)^2 + \beta \|\alpha(P(\mu)) - P(\mu) + Q\|_F^2 \right], & (b) \\ Q^{k+1} = Q^k + \alpha(P(\mu)) - P(\mu), & (c) \end{cases} \tag{10}$$

In the Bregman iteration, Eq. (10b) can be reduced to a $l_2$ minimization, which can be solved using the conjugate gradient (CG) method, which produces the exact solution after a finite number of iterations. The most difficult task is to solve the optimization (10a) because it contains differential of coordinate functions. Applying the standard variation method, Eq. (10a) is equivalent to solving following Laplace-Beltrami equation.

$$\begin{cases} -\Delta_M u(x) + \mu \sum_{y \in \Omega} \delta(x - y)(u(y) - v(y)) = 0, & x \in M \\ \frac{\partial u}{\partial n}(x) = 0, & x \in \partial M \end{cases} \tag{11}$$

where $n$ is the out normal of $M$, and $\partial M$ is the boundary of $M$. Recently, the point integral method has been proposed to solve Laplace-Beltrami equation over a point cloud [Ref]. The main idea of the point integral method is to apply following integral approximation for the differential term in Laplace-Beltrami equation:

$$\int_M \Delta_M u(y) R_t(x, y) dy \approx -\frac{1}{t} \int_M (u(x) - u(y)) R_t(x, y) dy + 2 \int_{\partial M} \frac{\partial u(y)}{\partial n} R_t(x, y) dy \tag{12}$$

where $R_t(x, y)$ are kernel functions given as follows,

$$R_t(x, y) = C_t \exp\left(-\frac{|x-y|^2}{4t}\right), \tag{13}$$

where $C_t$ is a normalizing factor. Using the integral approximation (12), following integral equation can be obtained to approximate the Laplace-Beltrami equation,

$$\int_M (u(x) - u(y))R_t(x, y) dy + \mu t \sum_{y \in \Omega} R_t(x, y)(u(x) - u(y)) = 0 \tag{14}$$

The integral equation (14) can be further discretized into a matrix equation over the point set $P(\mu)$ using some quadrature rule [12]:

$$\begin{cases} (L + \bar{\mu}W)U = \bar{\mu}WV \\ L = D - W \end{cases} \tag{15}$$

where $\bar{\mu} = \frac{\mu t N}{|M|}$, $\omega_{ij} = R_t(x_i, x_j), i, j = 1, 2, \cdots, N$, $W = (\omega_{ij} | i, j = 1, 2, \cdots, N)$ is the weight matrix, $d_i = \sum_{j=1}^{N} \omega_{ij}$ $D = diag(d_i | i = 1, 2, \cdots, N)$ is a diagnosis matrix. Thus, the optimization (10a) can be solved based on the matrix equation (15). The detailed formulation and alternating minimization steps for solving Eq. (10) are described in the flowchart for **Algorithm 1.**

**Algorithm 1**

---

*Initialize* an initial image $\mu$, $Q_0 = 0$ and parameters $\lambda$ and $\beta$;

1: **While** the current solution is not converged **do**

2: **Compute** the weight matrix $W = \{w_{i,j}\}$ from the patch image $P(\mu^{(n)})$, and the matrices

$L = D - W$, $D = diag(d_i)$, $d_i = \sum_j w_{i,j}$

3: **Solve** the linear systems: $\begin{cases} (L + \beta \cdot W)U = \beta \cdot WV \\ V = P(\mu^{(n)}) - Q^{(n)} \end{cases}$

4: **Update** $\mu$ by solving the problem: $\mu^{(n+1)} = \operatorname{argmin}\left[\frac{\lambda}{2}\sum_{i=1}^{N} b_i(A_i\mu - y_i)^2 + \beta\|U - P(\mu) + Q^{(n)}\|_F^2\right]$

5: **Update** $Q^{(n)}$: $Q^{(n+1)} = Q^{(n)} + U - P(\mu^{(n+1)})$

6: **End While**

---

## 3. Image reconstruction results

In the section, we test the proposed LDM-based reconstruction model with real patient datasets obtained on a GE clinical CT scanner. In addition, we also compare our results with those obtained by the conventional filtered back projection (FBP) method and the simultaneous algebraic reconstruction technique (SART) with a total variation (TV). All numerical computations in this section are implemented by MATLAB in a PC with 16G RAM and 2.8GHz CPU.

**3.1. Simulation result:** A realistic phantom adapted from a human CT slice is used to evaluate the proposed algorithms. We use an computer-assisted tomography simulation environment (CatSim) [17], which was developed by GE Global Research Center, to simulate x-ray imaging. CatSim incorporates polychromaticity, realistic quantum and electronic noise models, finite focal spot size and shape, finite detector cell size, and detector cross-talk for the simulation of real x-ray imaging. All acquisitions are simulated with polychromatic x-ray source operated at 120 kVp and 0.2mSv dose for the low dose imaging. The radius of the scanning trajectory is 54.1cm. 984 projections are uniformly acquired over a 360-degree angular range. For each projection, 888 detector elements are equiangular distributed. The phantom is discretized into a 512 × 512 matrix, and the sinogram is formed by stacking all projections of different views, as shown in Fig. 2. We perform image reconstruction respectively using the proposed LDM-based image reconstruction, the simultaneous algebraic reconstruction technique (SART) with a total variation (TV) and FBP method. The comparisons show that the LDM-based image reconstruction model is better than the other reconstruction method, as shown in Fig. 3.

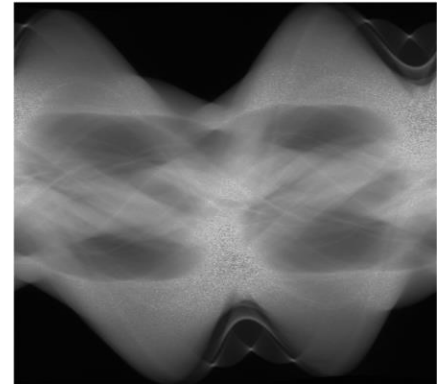

**Fig.2** The sinogram from CatSim.

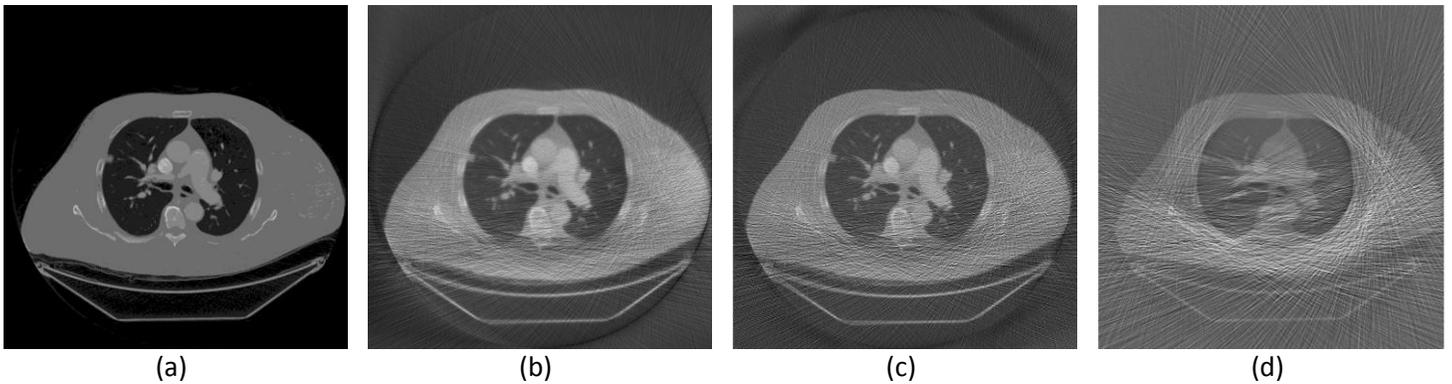

(a)  (b)  (c)  (d)

Fig. 3: Comparison of CT reconstruction results over lung tumor. (a) Ground truth CT images, (b) the reconstructed image using the proposed method; (c) the reconstructed image using SART with TV, and (d) the reconstructed image using a FBP method.

**3.2. Experimental results**

In this data set, the scan is in a typical helical geometry. After appropriate preprocessing, we obtained a set of 64-slice fan-beam sinograms, as shown in Fig. 3. The radius of the scanning trajectory is 54.1cm. Over a 360-degree angular range 984 projections are uniformly acquired. For each projection, 888 detector elements were equiangular distributed. The field of view (FOV) is of a 25 cm radius. The image matrix was of 512 × 512 pixels. Then, the sinogram is formed by stacking all projections of different views, as shown in Fig. 4.

From the sinogram, we first conduct image reconstruction using the proposed method. As the image illustrated in Fig.5 (a), the proposed LDM-based image reconstruction well preserves structural information especially texture features. In our method, we choose the patch size of 16x16 to form the patch manifold and regularization parameters are chosen as $\lambda = 0.5$ and $\beta = 0.2$. For comparison, the FBP method and the SART with TV regularization are applied as well to perform the image reconstruction from same projection dataset, whose results are showed in Fig .5 (b-c), respectively. The comparisons show that the LDM-based image reconstruction model outperforms the other two reconstruction methods. The image reconstructed via only SART iteration with TV is blurry. SART with TV is suitable to reconstruct simple structural images. For complex medical images, SART-TV over-smoothens textured regions, resulting in the loss of details. FBP keeps the structural information but it makes the reconstructed image noisy.

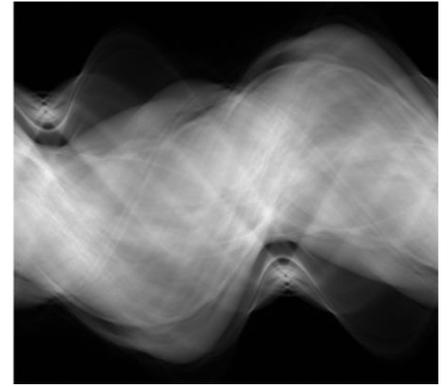

Fig. 4. The sinogram from a clinical scanner.

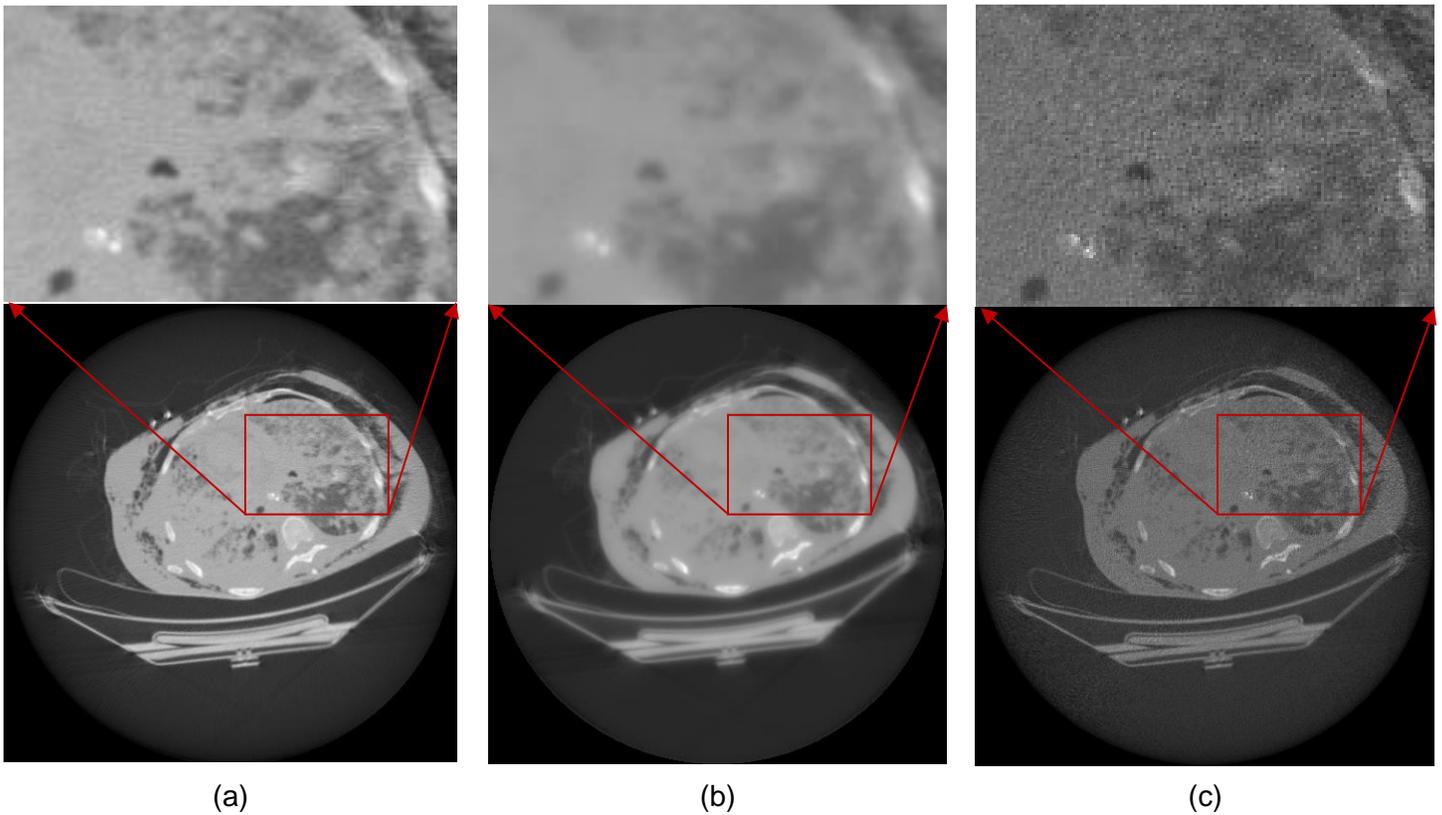

(a)                  (b)                  (c)

Fig. 5. Comparison of CT image reconstructions from raw patient data. (a) The reconstructed image using the LDM-based method, (b) the reconstructed image using SART with TV, and (c) the reconstructed image using FPB.

**4. Discussions and Conclusion**

The major contribution in this paper is to present an image reconstruction method aided by the regularization of a low dimensional manifold (LDM) model. This method promises substantially increased spatial and contrast resolution. Our iterative algorithm also incorporates prior knowledge, and account for photon statistics at a low dose level. However, the computational cost of the proposed LDM-based image reconstruction method is higher than the SART iterative methods. Major computational cost is matrix-vector multiplication operations in the iterative algorithm. This problem can be solved by parallel computation on GPU computer because matrix-vector multiplication is highly data parallel computation. The computational speed of the proposed iterative method can be improved on a GPU workstation.

The comparison between the proposed method and several representative methods has been performed to illustrate the merits of the LDMM-based reconstruction approach. The raw datasets from a clinical CT scanner have been used to evaluate the image quality. Results show that the regularization method of low dimensional manifold is an efficient and robust image reconstruction technique, and well preserves image edges and structural details of the reconstructed image comparing to the FBP method and the SART with TV regularization. This LDM-based approach is very promising for medical imaging and other applications.

**Acknowledgment**: This work is partially supported by the National Institutes of Health Grant NIH/NIBIB R01 EB016977 and U01 EB017140. R. Lai's work is partially supported by the National Science Foundation NSF DMS-1522645.